\documentclass[sigconf]{acmart}

\copyrightyear{2026}
\acmYear{2026}
\setcopyright{cc}
\setcctype{by}
\acmConference[MSR '26]{23rd International Conference on Mining Software Repositories}{April 13--14, 2026}
{Rio de Janeiro, Brazil}
\acmBooktitle{23rd International Conference on Mining Software Repositories (MSR '26), April 13--14, 2026,
Rio de Janeiro, Brazil}
\acmPrice{}
\acmDOI{10.1145/3793302.3793328}
\acmISBN{979-8-4007-2474-9/2026/04}

\usepackage[export]{adjustbox}
\usepackage{listings}
\usepackage{lstnix}
\usepackage{cleveref}
\usepackage{array}
\newcommand\LILA{Lila}

\author{Julien Malka}
\email{julien.malka@telecom-paris.fr}
\orcid{0009-0008-9845-6300}
\affiliation{\institution{LTCI, Télécom Paris, Institut Polytechnique de Paris}
	\city{Palaiseau}
	\country{France}
}

\author{Arnout Engelen}
\email{arnout@bzzt.net}

\begin{document}

\title{Lila: Decentralized Build Reproducibility Monitoring for the Functional Package Management Model}

\begin{abstract}

  Ensuring the integrity of software build artifacts is an increasingly important concern for modern software engineering, driven by increasingly sophisticated attacks on build systems, distribution channels, and development infrastructures. Reproducible builds---where binaries built independently from the same source code can be verified to be bit-for-bit identical to the distributed artifacts---provide a principled foundation for transparency and trust in software distribution.

  Despite their potential, the large-scale adoption of reproducible builds faces two significant challenges: achieving high reproducibility rates across vast software collections and establishing reproducibility monitoring infrastructure that can operate at very large scale. While recent studies have shown that high reproducibility rates are achievable at scale---demonstrated by the Nix ecosystem achieving over 90\% reproducibility on more than 80,000 packages---the problem of effective reproducibility monitoring remains largely unsolved.
  
In this work, we address the reproducibility monitoring challenge by introducing \textit{Lila}, a decentralized system for reproducibility assessment tailored to the functional package management model. \LILA{}~enables distributed reporting of build results and aggregation into a reproducibility database, benefiting both practitioners and future empirical build reproducibility studies.
\end{abstract}



\maketitle

\section{Introduction}

Ensuring the integrity of software build artifacts has become a central challenge in modern software engineering. Recent supply chain compromises---such as the SolarWinds breach and the backdoor inserted into the widely used \texttt{xz} compression tool---have highlighted the risks inherent in trusting opaque or unverified build artifacts. In response, governments and industry initiatives are promoting greater transparency and verifiability in software production and distribution processes.

Reproducible builds (R-B) provide a principled foundation for this transparency. By allowing independent parties to regenerate binaries from source and verify bitwise equivalence, reproducible builds shift assurance from trust in the distributor to trust in auditable and transparent build processes. This model complements cryptographic signing by verifying what was built, not just who built it. As a result, reproducible builds have emerged as a promising mechanism for increasing the integrity of software supply chains~\cite{lamb_reproducible_2022,malka_how_2025}. As a side effect, regularly checking software artifacts for bitwise reproducibility also ensures that software is \textit{rebuildable}, another very meaningful property for software engineering as rebuildability is a necessary condition for long term software preservation.

Despite their success in projects such as Debian or Arch Linux, practical adoption of reproducible builds still faces two major challenges.
First, there is doubt among experts on the feasibility of achieving high reproducibility rates across very large and diverse package collections~\cite{fourne_its_2023}. Encouragingly, a recent study has shown that the Nix package collection attains reproducibility rates exceeding 90\% across more than 80,000 packages, demonstrating that large-scale reproducibility is technically achievable~\cite{malka_does_2025}.
Second, and less explored, is the challenge of monitoring reproducibility at scale. Existing monitoring infrastructures are typically centralized, computationally expensive, and often limited to subsets of packages or recent versions.

In this work, we address the reproducibility monitoring challenge through \textit{Lila}, a decentralized system for assessing and aggregating reproducibility results in functional package management ecosystems. Lila enables distributed reporting of build outcomes from independent builders, which are then aggregated into a reproducibility database. This design enables scalable and transparent monitoring of reproducibility across large package collections. \LILA{} provides both a practical tool for developers and maintainers, and a foundation for future research on transparent, large-scale software build reproducibility.

\paragraph*{Paper structure}

We first introduce some background necessary to understand the design of our \LILA{}~tool in \Cref{sec:context}. We then present our \LILA{}~tool in \Cref{sec:lila}. Related work is presented in \Cref{sec:related}. \Cref{sec:discussion} presents future work opportunities and related discussions, and we conclude with \Cref{sec:conclusion}.

\section{Background}
\label{sec:context}

In this section, we provide the necessary context to understand the design of our tool and its impact.

\subsection{Reproducible builds}

Reproducible builds are a software engineering practice aimed at ensuring that a given source code, when built within a defined software environment, reliably produces bitwise identical binary artifacts—across different machines and over time. The primary goal of reproducible builds is to create a verifiable link between source code and its compiled output, enabling third parties to independently verify the integrity of distributed software. This approach directly addresses a fundamental trust assumption in the software supply chain: that published binaries faithfully correspond to their declared source code. A challenge to this approach is that traditional software build processes often produce non-identical outputs due to various sources of non-determinism. Such factors include embedded build timestamps, non-deterministic compiler behavior, differences in filesystem implementations, and other environment-specific variations.
The Reproducible Builds Project~\cite{noauthor_reproducible_2025}---launched in 2015 by members of the Debian community and later extended to various open-source ecosystems—has played a central role in defining the goals, methodologies, and best practices required to achieve build reproducibility. The project has provided tools, guidelines, and extensive documentation to help developers identify and eliminate sources of build non-determinism. Furthermore, its contributors have been instrumental in auditing the broader Free and Open Source Software (FOSS) ecosystem, proposing fixes for reproducibility issues, and working to have these fixes adopted upstream.

\subsection{Functional package management and Nix}

Nix~\cite{dolstra_purely_2006} and Guix~\cite{courtes_functional_2013} are implementations of the functional package management paradigm, introduced by Dolstra, which fundamentally differs from traditional package management models. In this approach, packages are treated as pure functions of their declared build- and run-time dependencies.
Nix expressions evaluate reproducibly into precisely defined build recipes, called \textit{derivations}, which specify the exact software environment required for the build~\cite{malka_reproducibility_2024}. 
Each derivation is uniquely identified by a cryptographic \textit{derivation hash}. When built, derivations produce \textit{outputs}---directories containing the resulting build artifacts---also uniquely identified by their \textit{output hash}, a value derived from the derivation metadata rather than the contents of the output itself. An example of such a build recipe for the \texttt{jq} package is shown in \Cref{fig:nix_expression}.

\paragraph*{Nixpkgs, the Nix Package Collection}

The Nix community maintains nixpkgs, a comprehensive collection of Nix expressions for building a vast range of software. Serving as the package repository for both the Nix ecosystem and the NixOS Linux distribution, nixpkgs currently contains over 80,000 packages~\footnote{The \href{https://repology.org/repositories/statistics}{repology} database reports about 85,000 packages as of July 2025.}. It aggregates software from various application domains and language-specific ecosystems, making it one of the largest package collections among Linux distributions.

\paragraph*{Store Substitution Mechanism}

Although nixpkgs is primarily a source-based distribution, Nix supports a binary substitution mechanism that allows users to replace locally built outputs with pre-built binaries from a binary cache. This allows users that wish to avoid heavy compilation to fall back to the more classical binary distribution model, but as a consequence requires them to use and trust a third-party binary distributor. Nixpkgs offers \texttt{cache.nixos.org}, a central binary cache containing all the packages built by the continuous integration platform of the project.

\begin{figure}[htbp]
\begin{adjustbox}{max width=\linewidth}
\begin{lstlisting}[language=nix]
{ lib, stdenv, fetchFromGitHub, autoreconfHook, bison}:

stdenv.mkDerivation rec {
  pname = "jq";
  version = "1.8.1";

  src = fetchFromGitHub {
    owner = "jqlang";
    repo = "jq";
    rev = version;
    hash = "sha256-R4ycoSn9LjRD/icxS0VeIR4NjGC8j/ffcDhz3u7lgMI=";
  };

  nativeBuildInputs = [ autoreconfHook bison ];
  configureFlags = "--with-oniguruma=no";

  preConfigure = ''
    echo "#!/bin/sh" > scripts/version
    echo "echo ${version}" >> scripts/version
    patchShebangs scripts/version
  '';

  meta = with lib; {
    description = "Lightweight command-line JSON processor";
    homepage = "https://jqlang.github.io/jq/";
    license = licenses.mit;
    platforms = platforms.unix;
  };
}
\end{lstlisting}
\end{adjustbox}
\caption{Simplified Nix expression for the \texttt{jq} package.}
\label{fig:nix_expression}
\end{figure}

\section{Lila}\label{sec:lila}

\begin{figure*}[ht]
  \includegraphics[width=0.60\textwidth, center]{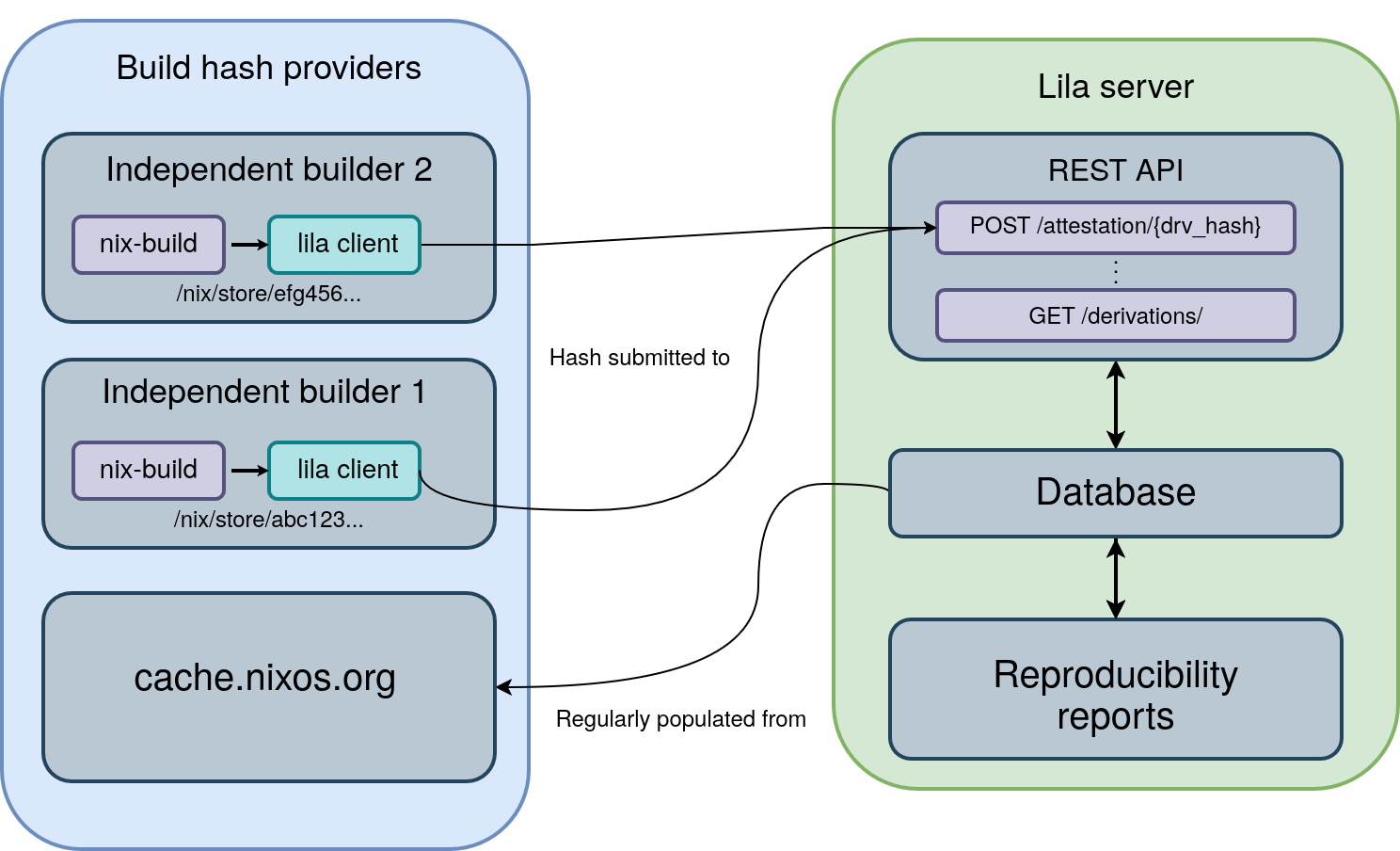}
  \caption{Overview of the Lila design.}
  \label{fig:lila-design}
\end{figure*}

An overview of \LILA{}'s design can be found in \Cref{fig:lila-design}.

\subsection{General design}

To enable decentralized collection of reproducibility attestations, the \LILA{}~client captures and reports build results directly from the Nix client-side build process. Its goal is to record, for each completed build, a verifiable statement binding the build recipe (derivation) to the produced outputs. To achieve this, \LILA{}~integrates with the Nix build system using the \texttt{post-build-hook} mechanism—a feature of Nix that allows the execution of arbitrary code after a local build finishes. Upon invocation, \LILA{}~inspects the output paths generated by the build and creates an attestation that includes the cryptographic hash of the derivation defining the build and the cryptographic hashes of the contents of all produced outputs. This attestation is signed with a builder-specific private key and submitted to the aggregation server through an authenticated API request.

The server component functions as a central aggregator for reported build hashes. It exposes a REST API (whose specification is detailed in \Cref{fig:rest-api}) which accepts signed attestations from authenticated clients. Clients authenticate using a private token that is attributed to them by the system administrator. This token is intended to avoid abuse of the aggregation server, but is not part of the trust chain between the creator and the consumer of the attestations. The server validates the reports and stores them for later querying and comparison.  Additionally, \LILA{} can ingest build hashes produced by the NixOS continuous integration platform, allowing the database to benefit from the comprehensive set of builds performed by that infrastructure.
An overview of the attestation structure is shown in~\Cref{fig:attestation_format}.

The server also includes a web interface designed to generate and display reproducibility dashboards and reports. This interface enables users to monitor sets of packages of interest and to easily detect reproducibility regressions over time.

\begin{table}
\begin{tabular}{|l|p{6cm}|p{5cm}}
\toprule
\textbf{Field} & \textbf{Description} \\
\midrule
\texttt{id} & Unique identifier for the attestation \\
\texttt{output\_path} & Store path of the built output \\
\texttt{user\_id} & Identifier of the submitting user \\
\texttt{drv\_id} & Identifier of the derivation that produced the output \\
\texttt{output\_hash} & Cryptographic hash of the output content \\
\texttt{output\_sig} & Signature of the hash by the submitting client \\
\bottomrule
\end{tabular}
\caption{Schema of an attestation entry in the \LILA{}~database.}
\label{fig:attestation_format}
\end{table}

\begin{table*}
\centering
\begin{tabular}{llp{8.5cm}}
\toprule
\textbf{Endpoint} & \textbf{Method} & \textbf{Description} \\
\midrule
\texttt{/attestation/\{drv\_hash\}} & POST & Submit a signed attestation for a derivation and its output hashes (requires token). \\
\texttt{/attestations/by-output/\{output\_path\}} & GET & Retrieve all attestations for a given output path. \\
\midrule
\texttt{/derivations/} & GET & List all derivations known to the server. \\
\texttt{/derivations/\{drv\_hash\}} & GET & Retrieve attestation summary or full attestation set for a specific derivation. \\
\midrule
\texttt{/reports} & GET & List the names of all defined reports. \\
\texttt{/reports/\{name\}} & GET & Return the queried report in a formatted output. \\
\texttt{/reports/\{name\}/suggested} & GET & Get a suggested list of derivations needing rebuilds, based on a report (requires token). \\
\bottomrule
\end{tabular}
\caption{Overview of the \LILA{}~REST API architecture.}
\label{fig:rest-api}
\end{table*}

\subsection{Potential usage}
\label{sec:usage}

\LILA{}~is designed as a foundational component for both practitioners and researchers seeking to advance the security of the software supply chain and study build reproducibility.

For researchers, \LILA{}~offers a valuable source of empirical data to better understand the challenges of build reproducibility at scale. Its decentralized architecture facilitates the collection of reproducibility reports from a diverse set of users operating across varied hardware and environments, making it particularly suitable for analyzing build reproducibility in specific software ecosystems. Furthermore, \LILA{}’s extensible design enables the integration of additional metadata collection mechanisms, which can help collect data specifically tailored to studying precise phenomena.
In line with this vision, metadata included in build attestations could be used to better understand the exact content of the build environment that was used to build a given derivation. Studying those metadata could help answer research questions until then difficult to tackle by academic research, such as the impact of bitwise reproducibility of each software component in the build environment on the bitwise reproducibility of the produced artifact. 

Beyond its research applications, \LILA{}~is intended as a practical tool for experts and maintainers working on reproducible builds. By providing reproducibility dashboards, \LILA{}~facilitates the timely identification of packages that have regressed in reproducibility, enabling faster response and remediation. Additionally, its decentralized model for distributing the verification workload expands both the scope and temporal reach of reproducibility monitoring within the NixOS community. This approach not only allows for monitoring a broader set of packages but also supports historical tracking—collecting build attestations for older package versions—thereby enabling long-term analysis of the reproducibility status of packages over time.

Finally, we hope that \LILA{}~can serve as a foundational building block for the development of user-facing tools that leverage bitwise reproducibility to minimize or eliminate trust relationships within the Nix binary distribution supply chain.

\section{Related work}
\label{sec:related}

This work is contributing to the literature on bitwise build reproducibility.

\paragraph*{R-B for software supply chain integrity}

Lamb et al. articulated a vision of a binary distribution system grounded in R-B, emphasizing its role in eliminating blind trust in distribution intermediaries~\cite{lamb_reproducible_2022}. However, some challenges are still open to achieve this vision and currently it remains largely unimplemented.

\paragraph*{Reproducibility monitoring systems}

Current reproducibility monitoring solutions are largely centralized and project-specific. For example, \texttt{rebuilderd} automates rebuilds and artifact verification within the Arch Linux ecosystem~\cite{the_rebuilderd_contributors_rebuilderd_2025}. Similarly, Debian and other distributions maintain in-house reproducibility check infrastructures. While effective within their scopes, these systems rely on centralized builders, inherently limiting their scalability.
In contrast, Lila aims to decentralize the attestation and monitoring process, thereby distributing both computational workload and trust---a design choice inspired by the limitations observed in centralized monitoring systems.

\paragraph*{Empirical studies on R-B}

Empirical research has provided valuable insights into the practical challenges of achieving reproducibility in large package sets. Malka et al.~\cite{malka_does_2025}~showed that over 90\% of the Nix package collection is reproducible, proving that achieving R-B at scale is possible. Other studies, such as Sharma et al.~\cite{sharma_canonicalization_2025}, examined reproducibility in specific ecosystems like Maven, highlighting ecosystem-specific reproducibility challenges. Finally, Benedetti et al.~studied reproducibility in six popular software ecosystems and reported on the variability of reproducibility rates across them and ways to improve it~\cite{benedetti_empirical_2025}. While insightful, these studies are limited in scope (of the package set studied or the time frame) because of their centralized nature. Lila complements these empirical efforts by proposing an infrastructure that can collect reproducibility data from decentralized sources, potentially serving both operational monitoring and future academic studies.

\section{Discussion and future work}
\label{sec:discussion}

\LILA{} is currently being gradually deployed within the Nix community, with the goal of eventually replacing the existing, limited reproducibility dashboards. Our experimental instance has already proved its usefulness, collecting more than 150,000 reproducibility reports and producing several reproducibility dashboards that have
helped experts identify reproducibility issues. The success of \LILA{}’s ambition to cover a scope as large as the entire Nix ecosystem depends on the willingness of trusted third parties to contribute computing resources for reproducibility monitoring. Encouragingly, several such participants have already expressed interest, giving us confidence in the viability of this approach. From a technical perspective, our initial deployment with a small group of users revealed no scalability issues inherent to the design.

As for future work, we intend to leverage the \LILA{}~database to perform large-scale studies on bitwise reproducibility in functional package manager ecosystems and its evolution over time. While \LILA{}~was initially developed with the Nix ecosystem in mind, its design is fully compatible with Guix and could even serve as a shared infrastructure between both ecosystems.

One limitation of \LILA{} is that it still relies on a centralized aggregation server for collecting and serving reproducibility attestations, and only decentralizes build execution and attestation across independent participants. Fully peer-to-peer aggregation would require substantially more engineering effort and introduce challenges related to coordination and consensus. We therefore view the current design as a pragmatic first step that can be used today to improve software supply chain security, while leaving more ambitious fully decentralized reproducibility tracking architectures for future work.

\section{Conclusion}
\label{sec:conclusion}

In this work, we presented \LILA{}, a decentralized infrastructure for monitoring build reproducibility within the functional package management ecosystem. By leveraging the properties of functional package managers like Nix, \LILA{}~enables scalable and decentralized collection of reproducibility attestations across diverse environments and participants. 

Our system complements existing centralized reproducibility monitoring solutions by distributing the verification effort, providing a persistent and queryable database of build reports, and allowing more secure package manager implementations to build upon that database. We envision \LILA{}~as a tool and a database to support future empirical research on build reproducibility, allowing the replication of existing methodologies at a scale not attainable before and answering new research questions.

\section{Availability}

\LILA{}~is open source software, its sources are released under the EUPL-1.2 licence~\cite{replication-code:1.0}.

\bibliographystyle{ACM-Reference-Format}
\bibliography{bibliography}
\end{document}